\documentclass[
aps,%
12pt,%
final,%
notitlepage,%
oneside,%
onecolumn,%
nobibnotes,%
nofootinbib,% 
superscriptaddress,%
noshowpacs,%
]{revtex4}
\usepackage{epsf,graphicx}
\newcommand{\be}{\begin{eqnarray}}
\newcommand{\ee}{\end{eqnarray}}

\def\nue{{\nu_e}}
\def\anue{{\bar\nu_e}}
\def\numu{{\nu_{\mu}}}
\def\anumu{{\bar\nu_{\mu}}}
\def\nutau{{\nu_{\tau}}}

\newcommand{\dm}{\mbox{$\Delta m_{21}^2$~}}

\newcommand{\kl}{\mbox{KamLAND~}}

\newcommand{\ms}{\Delta m^2_{21}}
\newcommand{\ma}{\Delta m^2_{31}}

\newcommand{\sss}{\sin^2 \theta_{12}}
\newcommand{\sch}{\sin^2 \theta_{13}}

\newcommand{\sa}{\sin^2 \theta_{23}}
\newcommand{\sta}{\sin^22 \theta_{23}}

\newcommand{\mat}{\Delta m^2_{31}{\mbox {(true)}}}

\newcommand{\scht}{\sin^2 \theta_{13}{\mbox {(true)}}}

\newcommand{\csh}{\sin^2 \theta_{13}}
\newcommand{\sat}{\sin^2 \theta_{23}{\mbox {(true)}}}

\newcommand{\pmm}{P_{\mu\mu}}
\newcommand{\sig}{$3\sigma$}
\newcommand{\evsq}{eV$^2$}
\newcommand{\obb}{0\mbox{$\nu\beta\beta$}}
\newcommand{\onbb}{neutrinoless double beta decay }
\newcommand{\meff}{\mbox{$\langle m \rangle$}}
\newcommand{\meffnh}{\mbox{$\langle m \rangle$}^{\rm NH}}
\newcommand{\meffih}{\mbox{$\langle m \rangle$}^{\rm IH}}

\newcommand{\meffihmin}{\mbox{$\langle m \rangle$}^{\rm IH}_{\rm min}}

\newcommand{\meffnhmax}{\mbox{$\langle m \rangle$}^{\rm NH}_{\rm max}}

\newcommand{\nme}{\mbox{nuclear matrix elements}}

\def\ltap{\ \raisebox{-.4ex}{\rlap{$\sim$}} \raisebox{.4ex}{$<$}\ }
\def\gtap{\ \raisebox{-.4ex}{\rlap{$\sim$}} \raisebox{.4ex}{$>$}\ }

\begin{document}

%\selectlanguage{russian}
%\selectlanguage{english} 

\title{Probing the neutrino mass matrix in next generation 
neutrino oscillation experiments}

\author{Sandhya Choubey}
\email{sandhya@thphys.ox.ac.uk}
\affiliation{
Rudolf Peierls Centre for Theoretical Physics,
University of Oxford, 1 Keble Road, Oxford OX1 3NP, UK
}

\begin{abstract}
We review the current status of the neutrino mass 
and mixing parameters needed to reconstruct the neutrino mass matrix.
A comparative study of the 
precision in the measurement of oscillation parameters 
expected from the next generation solar, atmospheric, 
reactor and accelerator based experiments is presented. 
We discuss the potential of $0\nu\beta\beta$ experiments in 
determining 
the neutrino mass hierarchy and the importance of 
a better $\theta_{12}$ measurement for it.
\end{abstract}

\maketitle

%%%%%%%%%%%%%%%%%%%%%%%%%%%%%%%%%%%%%%%%%%%%%%%%%%%%%%%%
\section{Introduction}
%%%%%%%%%%%%%%%%%%%%%%%%

The last few years have provided us with conclusive proof 
of the existence of oscillations and hence mass and 
mixing in the neutrino sector.
While the atmospheric neutrino data from Super-Kamiokande (SK)
and the accelerator data from the K2K have confirmed 
oscillations in the $\numu-\nutau$ sector with best-fit
$\ma=2.1\times 10^{-3}$ eV$^2$ and $\sin^22\theta_{23}=1$ 
\cite{skatm}, the combined data from the solar neutrino 
experiments and the latest spectacular results from the 
\kl reactor experiment can be explained only by oscillations
of $\nue(\anue)$ with best-fit $\ms=8.0\times 10^{-5}$ \evsq
and $\sss=0.31$ \cite{solar}.
Thus, having established the  
existence of neutrino mass and mixing, 
these results have proclaimed
a new era in Neutrino Physics, where the emphasis has shifted from
unveiling the reasons for solar/atmospheric neutrino deficit to
making increasingly precise measurements of neutrino oscillation
parameters, a 
prerequisite for any progress in our understanding of
the origin of the 
patterns of solar and atmospheric neutrino mass and mixing.

In the present article we discuss the 
possibilities of high precision
measurement of the solar and atmospheric neutrino 
oscillation parameters with future data from 
solar, reactor, atmospheric and 
long baseline neutrino experiments. We expound the 
possibility of measuring the deviation of $\theta_{23}$ 
from maximality using earth matter effects in atmospheric 
neutrinos. Finally we study the feasibility of using \obb{}
experiments to determine the neutrino mass hierarchy and point 
the necessity for better $\sss$ measurements for accomplishing
it.

%%%%%%%%%%%%%%%%%%%%%%%%%%%%%%%%%%%%%%%%%%%%%%%%%%%%%%%%
\section{Solar Neutrino Oscillation Parameters}
%%%%%%%%%%%%%%%%%%%%%%%%

%%%%%%%%%%%%%%%%%%%%%%%%%%%%%%%%%%%%%%%%%%%%%%%%%%%%%%%%
\subsection{Bounds from Current Solar and Reactor Experiments}
%%%%%%%%%%%%%%%%%%%%%%%%

%\begin{figure}
%\begin{center}
%\includegraphics[width=16.0cm, height=6cm]{sol_kl1_kl2.eps}
%\caption{\label{fig:currentsol}
%}
%\end{center}
%\end{figure}

\begin{table*}[htb]
\begin{tabular}{ccccc}
\hline
{Data set used} 
& Range of $\Delta m^2_{21}$ %[$10^{-5}$ eV$^2$]
& Spread in $\Delta m^2_{21}$ & $\sin^2\theta_{12}$
& Spread in $\sin^2\theta_{12}$ \cr
\hline\hline
{only solar} & $(3.3 - 18.4)$$\times 10^{-5}$ eV$^2$ &{69\%} 
& $0.24-0.41$ &26\%\cr
%{sol+162 Ty KL}& 4.9 - 10.7
%& 37\%
%& $ 0.21-0.39$ & 30\%  \cr
{solar + 766.3 Ty KL}& $(7.2 - 9.2)$$\times 10^{-5}$ eV$^2$ 
& 12\% & $0.25-0.39$ & 22\% \cr
\hline
{solar(SNO3) + 766.3 Ty KL} & $(7.2 - 9.2)$$\times 10^{-5}$ eV$^2$ 
& 12\%  & $0.26 - 0.37$ & 18\% \cr
{solar(SNO3) + 3KTy KL}  & $(7.6 - 8.6)$$\times 10^{-5}$ eV$^2$ 
& 6\%& $0.26 - 0.36$ & 16\% \cr \hline
\end{tabular} \\%[2pt]
\caption{
The 3$\sigma$ allowed ranges and \% spread  of $\Delta m^2_{21}$ and
$\sin^2\theta_{12}$ from a 1 parameter fit.
}
\label{tab1}
\end{table*}

We present in Table \ref{tab1} \cite{global}
the $3\sigma$ allowed range of 
$\ms$ and $\sss$ that we have from the current available solar 
and reactor neutrino data. We also show the 
corresponding ``spread'' 
%defined as 
\be
{\rm spread} = \frac{ prm_{max} - prm_{min}}
{prm_{max} + prm_{min}}\times 100~.
\label{error}
\ee
Sensitivity of \kl to the shape  
and hence distortion of the reactor $\anue$ 
induced positron spectrum, gives the experiment a 
tremendous ability to constrain $\ms$. However, 
we can see from Table \ref{tab1} that it is not much 
sensitive to the mixing angle $\theta_{12}$.
The average energy and distance in \kl corresponds to 
$\sin^2(\dm L/4E) \approx  0$, {\it i.e}, it is situated close to  
a Survival Probability MAXimum
(SPMAX).
This means that the coefficient
of the $\sin^22\theta_{12}$ term
in $P_{\bar{e}\bar{e}}^{KL}$ is relatively small,
smothering the sensitivity of  \kl  to
$\theta_{12}$ \cite{th12,shika}. 

Also given in Table \ref {tab1} are
the expected bounds on $\ms$ and $\sss$ after 
the prospective future data from the running SNO and 
\kl experiments become available. 
For SNO we assume that the third and final phase of the experiment will
measure the same Neutral Current (NC) and 
Charged Current (CC) rates as the salt phase, but with 
reduced errors of 6\% and 
5\% respectively. 
For \kl we simulate the 3 kTy data at $\ms=8.0\times 10^{-5}$ eV$^2$
and $\sss=0.3$ and use a systematic error of 5\%.
The uncertainty in $\ms$ is expected to reduce 
to 6\% with 3 kTy data from KamLAND. The uncertainty in 
$\sss$ is expected to improve after the phase-III results 
from SNO to 18\%. This would improve to about 16\% if the 
SNO phase-III projected results are combined with the 3 kTy 
simulated data from KamLAND. However, we 
note that even with the combined data from 
phase-III of SNO and 3 kTy statistics from KamLAND, the uncertainty 
on $\sss$ would stay well above the 10-15\% level at $3\sigma$.

%%%%%%%%%%%%%%%%%%%%%%%%%%%%%%%%%%%%%%%%%%%%%%%%%%%%%%%%
\subsection{Precision Expected from Next Generation Experiments}
%%%%%%%%%%%%%%%%%%%%%%%%%%%%%%%%%%%%%%%%

\begin{figure}[t]
\begin{center}
\includegraphics[width=14.0cm, height=7cm]{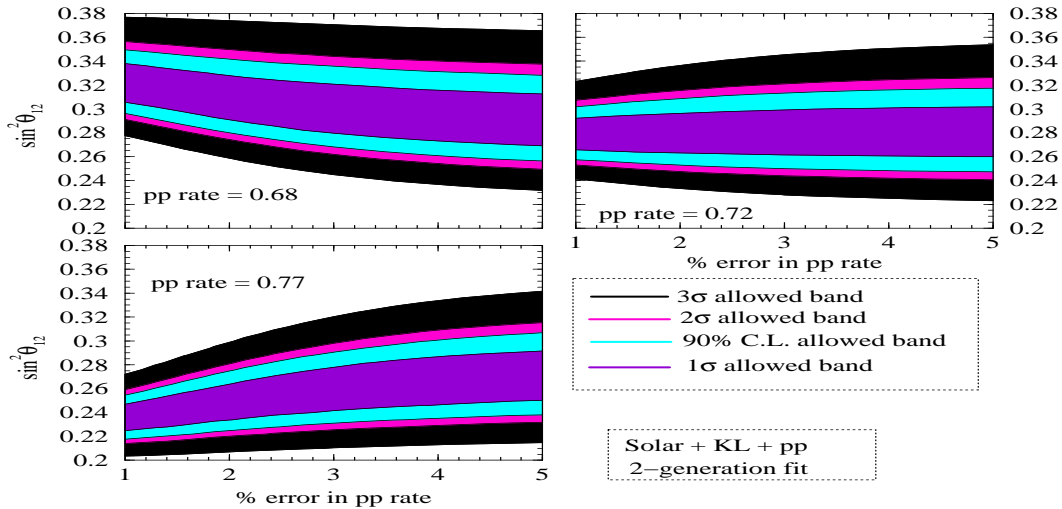}
\caption{\label{fig:pp}
Sensitivity plot showing the 
%1$\sigma$ (68.27\% C.L.), 1.64$\sigma$ (90\% C.L.),
%2$\sigma$ (95.45\% C.L.) and  3$\sigma$ (99.73\% C.L.)
allowed range of
$\sin^2\theta_{12}$ as a function of the error in $pp$ rate
for three different values of measured $pp$ rate.
%We also show the current 3$\sigma$ allowed range of $\sss$.
}
\end{center}
\end{figure}

\begin{figure}[t]
\begin{center}
\includegraphics[width=14.0cm, height=7cm]{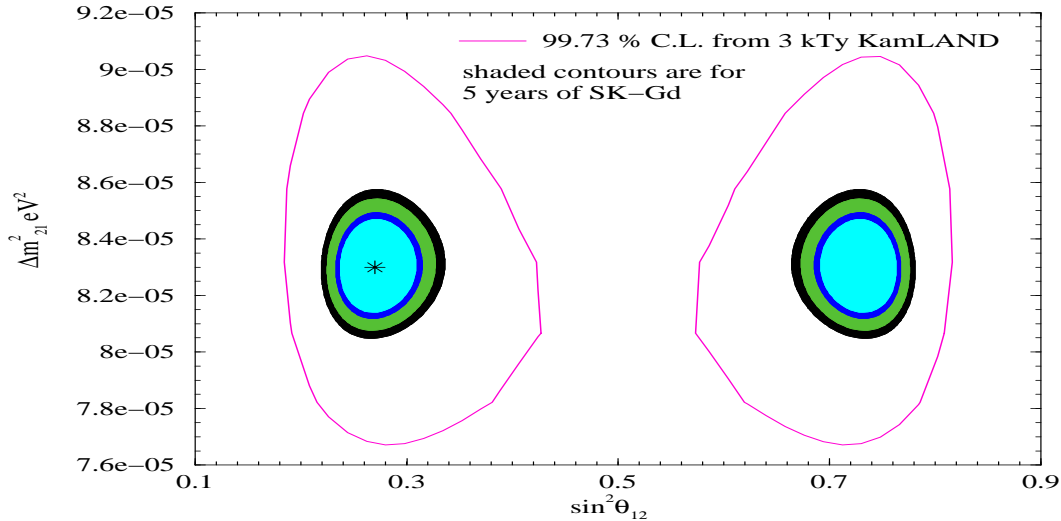}
\caption{\label{fig:skgd}
The 90\%, 95\%, 99\%, 99.73\% C.L. 
allowed regions in the $\ms-\sss$ plane from an
analysis of prospective data, 
obtained in 5 years of running of the SK-Gd detector.
The open contours shows the 99.73\% C.L. allowed areas 
expected from 3 kTy of \kl data. 
}
\end{center}
\end{figure}

In this sub-section we explore the possibility of new solar and 
reactor neutrino experiments and make a comparative study of  
their sensitivity to $\ms$ and
$\sss$. 
%This would give us a comparitive idea 
%of the strength of the different experiments as far as precision 
%measurements of the solar neutrino oscillation 
%parameters are concerned.

{\bf The Generic $pp$ experiment:}
Sub-MeV solar neutrino experiments (LowNu experiments) 
are being planned
for detecting the $pp$ neutrinos
using either charged
current reactions (LENS, MOON, SIREN) or
electron scattering process (XMASS, CLEAN, HERON, MUNU, GENIUS)
\cite{lownu}.  
It has been realized that high precision measurement of the 
$pp$ neutrino flux can be instrumental for  
more accurate determination of the neutrino mixing parameter, 
which as we have seen in the earlier section, will not be determined 
to an accuracy of below 10-15\% by the current set of experiments.  

We consider a
generic $\nu_e$-$e$ scattering experiment with a threshold of 50 keV.
This experiment will be sensitive to the
$pp$ neutrinos.
In Fig. \ref{fig:pp} we plot the two-generation
allowed range of $\sss$ from the 
global analysis of \kl and solar data including the LowNu $pp$ rate,
as a function of the error in the
$pp$ measurement.
We consider three illustrative $pp$ rates of 
0.68, 0.72 and 0.77 and
vary the experimental error in the $pp$ measurement from 1 to 5\%.
By adding the $pp$ flux data in the analysis,
the error in $\sin^2\theta_{12}$ determination
reduces to 14\% (19\%) at 3$\sigma$ for 1\% (3\%) uncertainty
in the measured $pp$ rate. Performing a similar 
three-neutrino oscillation analysis we have found that, 
as a consequence of the uncertainty on $\sin^2\theta_{13}$,
the error on the value of $\sin^2\theta_{12}$ increases 
correspondingly to 17\% (21\%) \cite{th12new}. 
%A LowNu measurement
%of the $pp$ neutrino flux with a 1\% error
%would allow to determine $\sin^2\theta_{12}$
%with an error of 14\% (17\%) at 3$\sigma$ 
%from a two-generation (three-generation)
%analysis. 

{\bf The SK-Gd reactor experiment:}
There has been a proposal to dope Super-Kamiokande (SK) with 
Gd by dissolving 0.2\% of gadolinium chloride in the SK water 
\cite{gdpaper}. SK gets the same 
reactor flux as KamLAND, and in principle could detect these reactor 
$\anue$ through their capture on protons, which 
releases a positron and a 
neutron. The detector has to tag these neutrons through delayed 
coincidence to be able to unambiguously observe the reactor $\anue$. 
However, neutron capture 
on proton releases a 2.2 MeV $\gamma$, which is undetectable in SK.
Addition of Gadolinium in SK would circumvent this problem since 
neutron capture on gadolinium releases a 8 MeV $\gamma$ cascade,
which is above the SK threshold and hence 
possible to observe. 
With its 22.5 kton of ultra pure water, the SK detector has 
about $1.5\times 10^{33}$ free protons as target 
for the antineutrinos coming 
from various reactors in Japan. 
Therefore for the same measurement period,
the SK-Gd reactor experiment is expected to 
have about 43 times the statistics 
of the \kl experiment.  

We simulate the reactor $\anue$ data expected 
%after 3 years of running of the 
in the proposed SK-Gd detector at 
$\ms=8.3\times 10^{-5}$ eV$^2$ and $\sss = 0.27$ and 
divide it into 18 energy bins, with a visible energy threshold
of 3 MeV and bin width of 0.5 MeV. The results of the
statistical analysis of this prospective data is presented  
in Fig. \ref{fig:skgd} for 5 years of exposure. 
Also shown in the figure for comparison is the 99.73\% 
C.L. line expected from a 3 kTy prospective data in KamLAND.
We can clearly see that the precision expected in both $\ms$ 
and $\sss$ is much better in SK-Gd. 
The spread in $\ms$ and $\sss$ 
expected from 5 years of data in SK-Gd
would be at the level of 2-3\% and 18\% respectively at \sig{}.
This can be compared with the corresponding spread of 
6\% and 32\% expected from 3 kTy of \kl data.
  
\begin{figure}[t]
\begin{center}
\includegraphics[width=14.0cm, height=7cm]{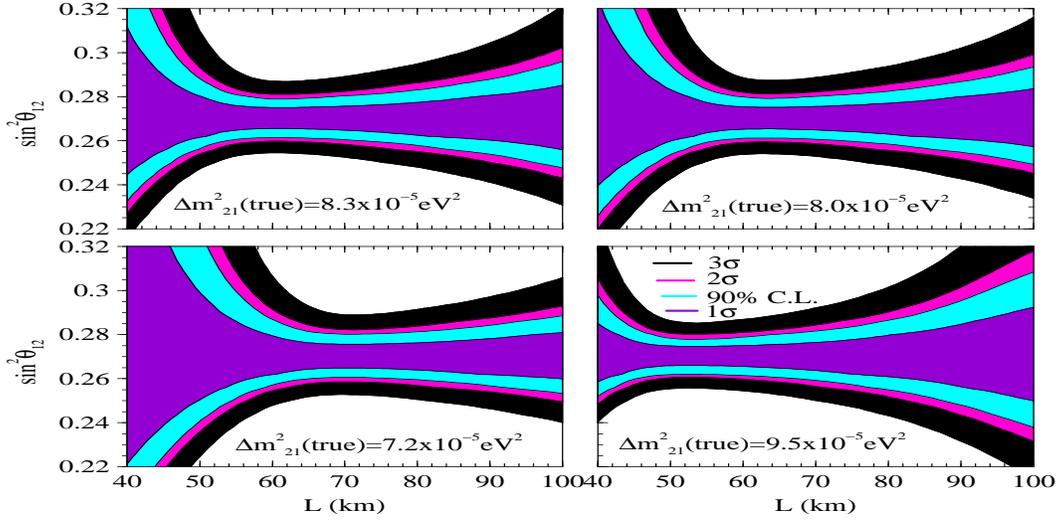}
\caption{\label{fig:spmin}
Sensitivity plots for the SPMIN reactor experiment
showing the $1\sigma$,  $1.64\sigma$, $2\sigma$, 
and $3\sigma$ range of allowed values for $\sss$ 
as a function of the baseline $L$. 
}
\end{center}
\end{figure}

{\bf The SPMIN reactor experiment:}
The solar mixing angle could be measured with unprecedented accuracy 
in a reactor experiment with the baseline tuned to the 
Survival Probability MINimum (SPMIN). This idea was 
proposed in \cite{th12}, where the optimal baseline 
for the most accurate measurement of $\sss$ for the old 
low-LMA best-fit $\ms=7.2\times 10^{-5}$ eV$2$ was found to be 70 km
from a statistical analysis. 
%Here we make a detail study of the 
%dependence of the spread in $\sss$ on
%the true value of $\ms$,
%the baseline, 
%the statistics and the systematics of the experiment. 
%We also compare the results of the two-generation analyses with a full 
%three-generation calculation, where the unknown parameter 
%$\sch$ is allowed to vary freely within its current allowed 
%range. 

In Fig. \ref{fig:spmin}
we show the $\sss$ sensitivity expected 
in a reactor experiment as a function of the baseline $L$.
We assume a total systematic uncertainty of
2\% and consider a statistics of 73 GWkTy (given as a product of 
reactor power in GW and the exposure of the detector in kTy).
We assume that the true value of $\sss=0.27$ and simulate the 
prospective observed positron spectrum in the detector for 
four different assumed true values of $\ms$. 
We simulate the spectrum at each baseline and plot the 
range of values of $\sss$ allowed by the experiment as a function 
of the baseline. 
The baseline at which the band of allowed values of $\sss$ is 
most narrow is the ideal baseline for the SPMIN reactor experiment.
The figure confirms that this ideal baseline depends crucially
on the true value of $\ms$. 
The most optimal baseline for 
the true value of 
$\ms=8.3(8.0)\times 10^{-5}$ eV$^2$ is 
seen to be $60 (63)$ km. 
At the most optimal baseline, the SPMIN 
reactor experiment can achieve an unprecedented
accuracy of $\sim 2(6)\%$ at $1\sigma(3\sigma)$ in 
the measurement of $\sss$.

From the Fig. \ref{fig:spmin} we get an impression that the 
optimal baseline for a given true value of $\ms$ is very 
finely tuned. 
However, note that in Fig. \ref{fig:spmin} we had allowed 
$\ms$ to vary freely. 
The uncertainty in the 
$\ms$ measurement translates to extra uncertainty in the $\sss$ 
measurement. If $\ms$ could be fixed, assumed to be measured to 
a very high precision in some other experiment like \kl or SK-Gd, 
then the uncertainty 
in $\sss$ due to $\ms$ can be drastically reduced.
If $\ms$ was kept fixed, 
then the choice of the baseline for setting up 
the SPMIN experiment becomes much broader \cite{th12new}. 

One of the driving features for this kind of a precision experiment 
is the statistics. 
We have checked that the sensitivity of 
$\sss$ improves from 3(10)\% to 2(6)\% at $1\sigma(3\sigma)$ as 
the statistics is increased from 20 GWkTy to 60 GWkTy. 
Another important aspect is the systematic 
uncertainty. 
Fig. \ref{fig:spmin} 
has been generated with an assumed 2\% systematic uncertainty,
which could be experimentally very challenging.
The effect of systematics on the $\sss$ 
measurement can be checked by repeating the
analysis with a more conservative 
estimate of 5\% for the systematic uncertainty. We find that for
$\ms(true)=8.3 \times 10^{-5}$ eV$^2$,
the spread in $\sss$ at $L=60$ km
increases from 6.1\% to 8.6\% at $3\sigma$, as the systematic error is 
increased from 2\% to 5\%. 
Finally, the impact of the $\theta_{13}$ uncertainty 
on the precision of $\sss$ is to increase the uncertainty in $\sss$ 
from 6.1\% to 8.7\% at $3\sigma$, for 
$\ms(true)=8.3 \times 10^{-5}$ eV$^2$ and $L=60$ km.

%%%%%%%%%%%%%%%%%%%%%%%%%%%%%%%%%%%%%%%%%%%%%%%%%%%%%%%%
\section{Atmospheric Neutrino Oscillation Parameters}
%%%%%%%%%%%%%%%%%%%%%%%%

%%%%%%%%%%%%%%%%%%%%%%%%%%%%%%%%%%%%%%%%%%%%%%%%%%%%%%%%
%\subsection{Current status}
%%%%%%%%%%%%%%%%%%%%%%%%

The parameters $\ma$ and $\sa$ at present 
are best constrained by the SK atmospheric neutrino data.
The \sig{} allowed ranges (spread) of $\ma$ and $\sa$ obtained from 
the final analysis by the SK collaboration are
(1.3-4.2)$\times 10^{-3}$eV$^2$ (53\%) and 0.33-0.66 (34\%) 
\cite{skatm}. The addition of the K2K data sample into the 
analysis brings in a modest improvement in the range of $\ma$ 
to (1.4-3.3)$\times 10^{-3}$eV$^2$ (42\%), however the 
the uncertainty in the allowed range of $\sa$ remains the same 
\cite{maltoniglobal}.

%%%%%%%%%%%%%%%%%%%%%%%%%%%%%%%%%%%%%%%%%%%%%%%%%%%%%%%%
%\subsection{Precision Expected in the Future}
%%%%%%%%%%%%%%%%%%%%%%%%

Both $\ma$ and $\sta$ are expected to be measured very 
accurately by the forthcoming long baseline (LBL) experiments.
A statistical analysis of the combined data set with
five years of 
running of MINOS, ICARUS, OPERA, T2K and NO$\nu$A 
each, reveals that $\ma$ and $\sa$ could be measured with 
a spread of 4.5\% 
and 20\% respectively at \sig{} \cite{huber10}.
The future prospective data from water Cerenkov atmospheric 
neutrino experiments with a statistics 20 times the current 
SK statistics could measure $\ma$ and $\sa$ with a spread of 
$\sim 17\%$ and $\sim 24\%$ respectively \cite{maltonimax23}.
A large magnetized iron calorimetric detector
such as the proposed
INO detector ICAL \cite{ino}, could use atmospheric neutrinos to
measure $\ma$ 
and $\sa$ with an 
accuracy comparable to that expected from the combined 
data from the LBL experiments
%of 3\% and 18\% respectively at \sig{} 
\cite{inootherparams}.

%%%%%%%%%%%%%%%%%%%%%%%%%%%%%%%%%%%%%%%%%%%%%%%%%%%%%%%%
\subsection{Is the Mixing Angle $\theta_{23}$ Maximal?}
%%%%%%%%%%%%%%%%%%%%%%%%

%%%%%%%%%%%%%%%%%%%%%%%%%%%%%%%%%%%%%%%%%
%\begin{figure}
%\begin{center}
%\includegraphics[width=14.0cm, height=9cm]
%{udEbins_d31=2.0_s13sq_0.0257_resol.eps}
%\caption{
%The up-down asymmetry expected for muon neutrinos in energy 
%bins of width 2 GeV. 
%The six panels show the data for 
%six different zenith angle
%(or $L$) 
%bins corresponding to upward neutrinos travelling 
%between $L_m=0-2000$ km, $L_m=2000-4000$ km, $L_m=4000-6000$ km,
%$L_m=6000-8000$ km, $L_m=8000-10000$ km and 
%$L_m=10000-12000$ km respectively inside the earth, cf. Table 
%\ref{tab:zen}.
%The solid black lines and the solid magenta lines are for 
%neutrinos travelling in matter with $\sa=0.5$ and 0.36 respectively.
%The dashed black lines and the dashed magenta lines are for 
%neutrinos travelling in vacuum with $\sa=0.5$ and 0.36 respectively.
%For all cases we have taken $\ma=2\times 10^{-3}$ eV$^2$ and 
%the benchmark 
%values of Table \ref{tab:benchmark} for the other parameters.
%}
%\label{fig:updown}
%\end{center}
%\end{figure}
%%%%%%%%%%%%%%%%%%%%
%%%%%%%%%%%%%%%%%%%%%%%%%%%%%%%%%%%%%%%%%
\begin{figure}
\begin{center}
\includegraphics[width=14.0cm, height=7cm]
{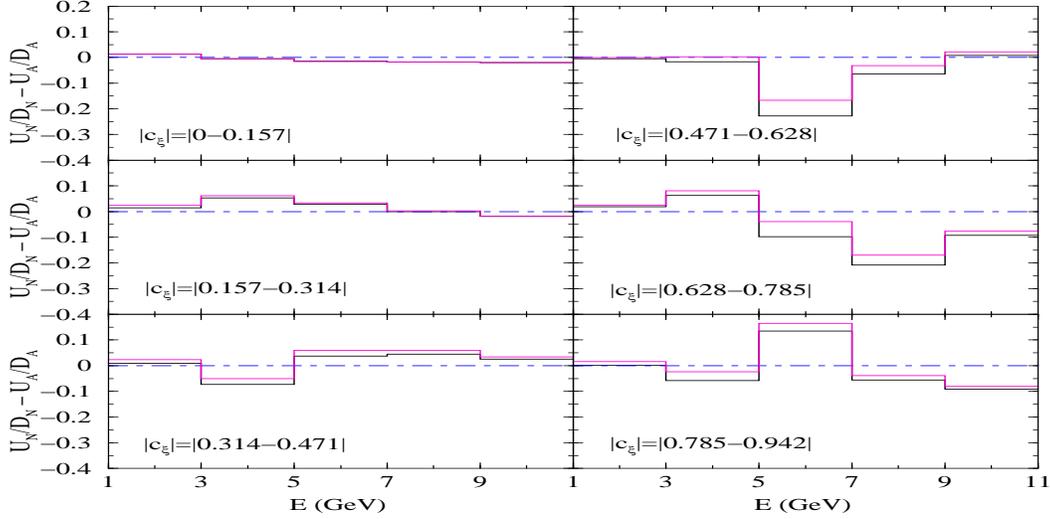}
\caption{
The difference between the 
up-down ratio for the neutrinos ($U_N/D_N$) and 
antineutrinos ($U_A/D_A$)
shown for the various energy and zenith angle bins. 
The solid black and solid magenta
lines are for neutrinos/antineutrinos 
travelling in matter 
with $\sa=0.5$ and 0.36 respectively.
%The dot-dashed blue line 
%shows $U_N/D_N - U_A/D_A =0$ for reference.
%The other 
%oscillation parameters are chosen as in Fig. \ref{fig:updown}.
}
\label{fig:updowndiff}
\end{center}
\end{figure}
%%%%%%%%%%%%%%%%%%%%

%%%%%%%%%%%%%%%%%%%%%%%%%%%%%%%%%%%%%%%%%
\begin{figure}[t]
\begin{center}
\includegraphics[width=14.0cm, height=7cm]
{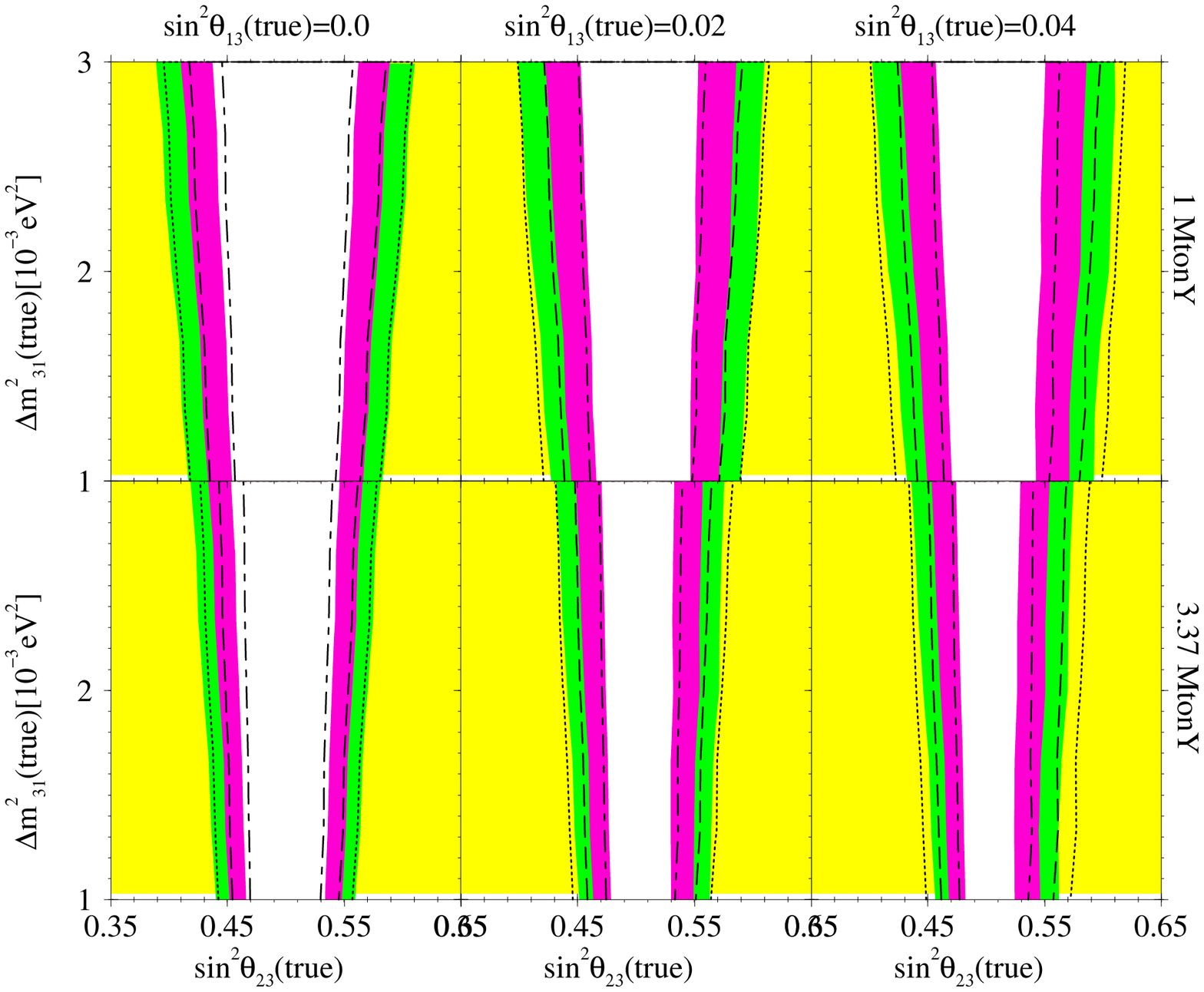}
\caption{
The regions of $\mat$ and $\sat$ where maximal $\theta_{23}$
mixing can be rejected by using 1 MtonY (upper panels) 
and 3.37 MtonY (lower panels) atmospheric neutrino 
data in ICAL
at $1\sigma$ (white band), $2\sigma$ (blue band)
and $3\sigma$ (green band). The hollow dark lines show the 
corresponding bands for neutrinos travelling in pure vacuum.  
%Additional constraint on $\sch$ 
%coming from combined data from future reactor and accelerator 
%data is taken into account. 
%The left, middle and right panels are for $\sch{\rm (true)}=0.0$,
%0.02 and 0.04 respectively. 
}
\label{fig:sens}
\end{center}
\end{figure}
%%%%%%%%%%%%%%%%%%%%

The measurement of both the magnitude and sign of   
the deviation of $\sa$ from its maximal value 0.5 is of 
utmost theoretical importance. In this section we will 
see how well the next generation of experiments can answer the 
question: Is the mixing angle $\theta_{23}$ maximal? 
To quantify the deviation of the true value of $\theta_{23}$ from 
its maximal value, we introduce the function
%\be
$D \equiv \frac{1}{2} - \sa.$
%\label{Eq:D}
%\ee
The best current limit on 
$D$ comes from the SK atmospheric neutrino 
experiment giving $|D| \leq 0.16$ at \sig{} \cite{skatm}.
More precise measurements of $\theta_{23}$ 
%and hence of $D$ 
are expected from 
future high statistics 
atmospheric neutrino data 
in water Cerenkov detectors. 
Better constraints are expected 
from forthcoming LBL experiments,
owing to their larger statistics and lower systematic errors.
However, in both these classes of experiments,
$\theta_{23}$ will be mainly determined by the 
$\numu$ (and/or $\anumu$) disappearance channel in vacuum,
which predominantly 
depends on $\sin^22\theta_{23}$. This leads to two very important 
consequences. Firstly, 
the fact that $\theta_{23}$ is close to maximal and 
$\delta(\sa) = \delta(\sin^22\theta_{23})/(4~\cos2\theta_{23})$ 
implies that even though one 
could determine the value of
$\sin^22\theta_{23}$ at the percentage level from the next generation 
long baseline experiments, 
the uncertainty 
in $\sa$ would still remain in the region of 10-20\%, depending 
on the true value of $\sa$ ($\sat$). 
The second consequence of the predominant 
$\sin^22\theta_{23}$ dependence of $P_{\mu\mu}$
means that they are almost insensitive 
to the octant of $\theta_{23}$ and hence to the sign of $D$. 

Large matter effects are known to exist in $\pmm$ for 
neutrinos travelling over very long distances with 
$L \sim 7000$ km. However, the net matter effects given 
by $\Delta \pmm (=\pmm - \pmm^{vacuum})$ changes sign 
with energy $E$ and baseline length $L$. Therefore, in order 
to maximize the observation of matter effects in the 
atmospheric neutrino disappearance probability, one needs a 
detector with very good energy and zenith angle ($\xi$) resolution.
Large magnetized iron calorimeter detectors like the one 
proposed for the ICAL detector at INO \cite{ino}, 
allows for good 
$E$ and $L$ resolution and at the same time can distinguish the 
$\numu$ from the $\anumu$ signal, 
further enhancing the sensitivity to
matter effects. Observation of such large matter effects in 
atmospheric neutrinos can be used for unambiguous determination
of the neutrino mass hierarchy \cite{inohier}. It has been 
quantitatively argued in \cite{inomax} that 
presence of large matter effects in $\pmm$ enhances its sensitivity
to $\theta_{23}$ as well.\footnote{See \cite{minosmax} for a 
discussion on using matter effects in $\pmm$ with $L\sim 1000$ km
to test maximality.}
Fig. \ref{fig:updowndiff} shows the impact of matter 
effects on atmospheric neutrinos events in an ICAL-like 
detector, for the 
normal mass hierarchy.
%\footnote{In this and the next subsection we assume
%the normal mass hierarchy for neutrinos}.
The figure shows the difference in the up-down ratio ($U/D$)
of the $\numu$ ($N$) and $\anumu$ ($A$) 
events, and have been binned 
into five energy and 12 zenith angle bins. 
For instance, the first (last) panel on the top left
(bottom right) contains $U_N/D_N - U_A/D_A$ for 
$0 \leq |\cos\xi| \leq 0.157$ ($0.785 \leq |\cos\xi| \leq 1$)
and calculated by taking upward going neutrinos with 
$-0.157 \leq \cos\xi \leq 0$ ($-1\leq \cos\xi \leq -0.785$)
and downward going neutrinos with 
$0 \leq \cos\xi \leq 0.157$ ($0.785\leq \cos\xi \leq 1 $).
We note that matter effects are largest in 
the $E=5-7$ GeV and $-0.628 \leq \cos\xi \leq -0.471$ 
bin. We also note that reducing $\sa$ from the maximal 
0.5 to 0.36 brings a nearly 
10\% change in the difference   $U_N/D_N - U_A/D_A$.

We perform a statistical analysis of the simulated data sample
in an ICAL-like detector, adding
the constraint on $\sin^2\theta_{13}$ expected from future reactor
and accelerator experiments \cite{huber10}.  
Our procedure is to generate the data at
a certain nonmaximal value of $\sin^2\theta_{23}$(true) and then fit
this data with the maximal $\sa = 0.5$,
choosing different values of $\sin^2\theta_{23}$(true). 
The details of 
the statistical analysis can be found in \cite{inomax}. 
The results are displayed in Fig. \ref{fig:sens}, where
the upper and lower panels
correspond to respective exposures of 1 MtonY and 
3.37 MtonY.
%\footnote{3.37 MtonY would
%roughly be the statistics needed in ICAL to match the number of
%fully contained muon events in the experiment SK20 which is
%described later in this section.}.
The regions of $\sin^2\theta_{23}$(true) and $\Delta m^2_{31}$(true)
within the white, blue and green bands of Fig. \ref{fig:sens} 
show the true values
of those quantities for which the distinction of a maximal from a true
nonmaximal value of $\theta_{23}$ will not be possible at the
$1\sigma$, $2\sigma$ and $3\sigma$ levels respectively for the
specified exposure.  The broken lines give the corresponding limits of
$\sin^2\theta_{23}$(true) in case earth matter effects were
deliberately switched off by hand.  
%Among those the long-dashed,
%dot-dashed and dotted lines respectively yield the $1\sigma$,
%$2\sigma$ and $3\sigma$ limiting values of $\sin^2\theta_{23}$(true).  
A comparison of the broken lines with the corresponding
continuous lines show that matter effects tend to
increase somewhat the sensitivity of ICAL to test the maximality of
$\sin^2\theta_{23}$. 
Specifically, for $\Delta m^2_{31}$(true) $= 2.0 \times 10^{-3} \
{\rm eV}^2$ and $\sin^2\theta_{13}$(true) $=0.04 ~(0.00), \
\sin^2\theta_{23}$(true) can be distinguished by ICAL from 
the maximal value of 0.5
at the $3\sigma$ level 
%for $\sin^2 \theta-{23}$(true) $= 0.420~(0.405)$, i.e. 
within 17\% (20\%) from 1 MtonY of exposure and 
%for $\sin^2\theta_{23}$(true) $= 0.443 ~(0.428)$, i.e. 
within 11\% (14\%) if the statistics was increased to 3.37 MtonY. 
This is comparable to the sensitivity of 
the combined data from the 
forthcoming accelerator-based LBL experiments to a
deviation from maximality of $\sin^2\theta_{23}$, which is 
\cite{antusch} $\sim
14$\% at $3\sigma$ 
(for $\Delta m^2_{31}$(true)$ = 2.5\times 10^{-3}$ eV$^2$).
Our sensitivity to $D$ is
also comparable to that expected with atmospheric neutrinos in 
very large futuristic water
Cerenkov detectors.  For statistics that is 20 (50) times the
current SK statistics, denoted as SK20 (SK50), a very large water
Cerenkov atmospheric neutrino experiment 
is expected to test a deviation from a maximal
$\sin^2\theta_{23}$ upto \cite{maltonimax23} 23\% (19\%) at $3\sigma$.

%%%%%%%%%%%%%%%%%%%%%%%%%%%%%%%%%%%%%%%%%%%%%%%%%%%%%%%%%%%%%%%%
\subsection{\label{sec:octant}
Sensitivity to the octant of $\theta_{23}$}
%%%%%%%%%%%%%%%%%%%%%%%%%%%%%%%%%%%%%%%%%%%%%%%%

%%%%%%%%%%%%%%%%%%%%%%%%%%%%%%%%%%%%%%%%%
\begin{figure}
\begin{center}
\includegraphics[width=14.0cm, height=7.0cm]
{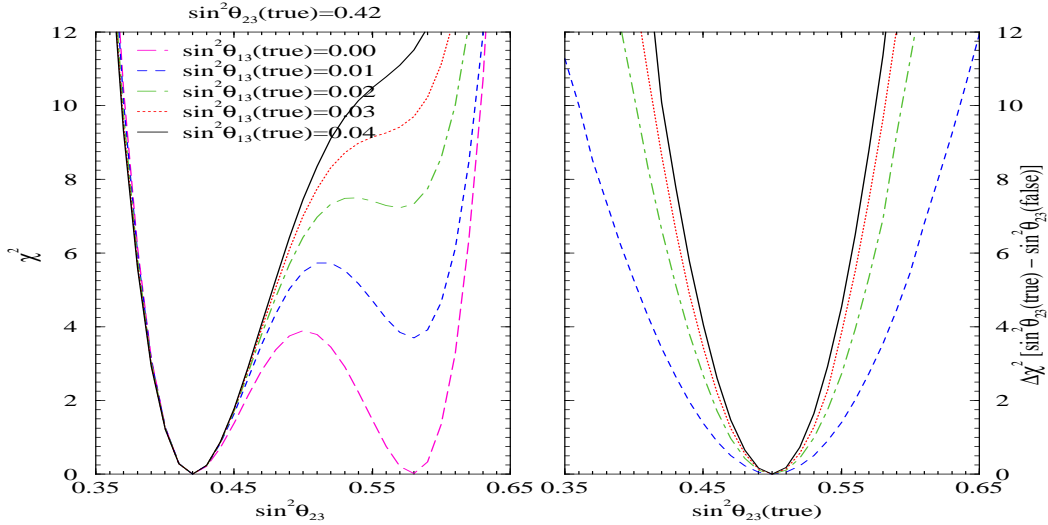}
\caption{Plot showing the octant sensitivity of ICAL.
Left-hand panel shows
$\chi^2$ as a function of $\sa$ for $\sat=0.42$. 
Right-hand panel gives the $\Delta \chi^2$ 
as a function of $\sat$.
}
\label{fig:delchioctant}
\end{center}
\end{figure}
%%%%%%%%%%%%%%%%%%%%
%\begin{figure}[t]
%\begin{center}
%\includegraphics[width=14.0cm, height=7.0cm, angle=0]
%{delchisq_resol_octant_true.eps}
%\caption{
%$\Delta \chi^2$ as a function of $\sat$ showing the 
%octant sensitivity of ICAL for the normal (left panel) and 
%inverted (right panel) neutrino mass ordering.
% The magenta long-dashed lines, blue short-dashed lines,
%green dot-dashed lines, red dotted lines and black solid lines
%are for $\sch{\rm (true)}=0.00$, 0.01, 0.02, 0.03 and 0.04 respectively.
%}
%\label{fig:delchioctanttrue}
%\end{center}
%\end{figure}
%%%%%%%%%%%%%%%%%%%%

If the true value of $\theta_{23}$ is not $45^\circ$, then 
the question arises whether $\theta_{23} >$ ($D$ positive)  
or $\theta_{23} < \pi/4$ ($D$ negative). This leads to an 
additional two-fold degeneracy in the 
measurement of the mixing angle $\theta_{13}$ and the CP phase 
$\delta$ in LBL experiments. This ambiguity is generally 
regarded as the most difficult to resolve.
As discussed before, the SK $\numu$/$\anumu$ data
is insensitive to the sign of $D$. The measurements from the 
disappearance channel in the forthcoming LBL experiments are 
also going to be insensitive to the sign of $D$. 
However, matter effects in $\pmm$ 
open up a new possibility for an ICAL-like detector -- sensitivity
to the octant of $\theta_{23}$.

Fig. \ref{fig:delchioctant} shows the results 
of our statistical analysis based on simulated data from 1 MtonY 
exposure in an ICAL-like atmospheric $\numu/\anumu$ experiment. 
The left-hand panel gives the $\chi^2$ as a function of $\sa$, 
for a ``data'' simulated by assuming that $\sat=0.42$.
For every nonmaximal
$\sin^2\theta_{23}$(true), there exists a $\sin^2\theta_{23}$(false)
which is given by 
$\sin^2\theta_{23}({\rm false}) = 1 - \sin^2 \theta_{23}({\rm true})$
on the other side of $\pi/4$.  
For a vanishing $\sin^2\theta_{13}$(true)
there are no matter effects and the $\chi^2$ corresponding to 
both the true and false values
of $\sin^2 \theta_{23}$ are the same. Hence they are 
allowed at the same C.L. and
one fails to fix the octant of $\theta_{23}$ in this case.
However, for
$\scht \neq 0$, matter effects bring in an octant
sensitivity and a false $\sin^2\theta_{23}$ solution can be ruled out,
provided $D$(true) is not too close to zero. 

In order to obtain the limiting value of $\sat$ 
which could still allow for the determination of $sgn(D)$
we define
\be
\Delta \chi^2 \equiv \chi^2 (\sin^2\theta_{23} ({\rm true}),
\sin^2\theta_{13} ({\rm true}), {\rm others(true)}) 
%\nonumber \\[2mm] 
%&& \hspace*{1cm} 
- \chi^2(\sin^2\theta_{23} ({\rm
false}),\sin^2\theta_{13}, {\rm others}),
\label{Eq:chioctant}
\ee
with `others' comprising $\Delta m^2_{31}$, $\Delta m^2_{21}$,
$\sin^2\theta_{12}$ and $\delta$.  These, along with
$\sin^2\theta_{13}$, are allowed to vary freely in the fit.  
The right-hand panel of Fig. \ref{fig:delchioctant} 
shows $\Delta\chi^2$ as a function of 
$\sin^2\theta_{23}$(true) for different values of $\scht$. 
The range of
$\sin^2\theta_{23}$(true), for which $\sin^2 \theta_{23}$(false) can
be ruled out 
is visible from the figure.  
In particular, 
%for a normal neutrino mass ordering,
$\sin^2\theta_{23}$(false) should be excludable at the
$3\sigma$ level from 1 MtonY of ICAL exposure for
%\be
%\sin^2\theta_{23} ({\rm true}) < 0.366 \ {\rm or} \ > 0.627 \ {\rm
%for} \ \sin^2\theta_{13} ({\rm true}) = 0.01, \\[2mm]
%\sin^2\theta_{23} ({\rm true}) < 0.405 \ {\rm or} \ > 0.588 \ {\rm
%for} \ \sin^2\theta_{13} ({\rm true}) = 0.02, \\[2mm]
%\sin^2\theta_{23} ({\rm true}) < 0.418 \ {\rm or} \ > 0.576 \ {\rm
%for} \ \sin^2\theta_{13} ({\rm true}) = 0.03,\\[2mm]
$\sin^2\theta_{23} ({\rm true}) < 0.424 \ {\rm or} \ > 0.570 \ {\rm
for} \ \sin^2\theta_{13} ({\rm true}) = 0.04$.
%\ee

%%%%%%%%%%%%%%%%%%%%%%%%%%%%%%%%%%%%%%%%%%%%%%%%%%%%%%%%
\section{Neutrino Mass Hierarchy from Future $0\nu\beta\beta$ Experiments}  
%%%%%%%%%%%%%%%%%%%%%%%%

%\begin{figure}
%\hglue -8.0cm
%\includegraphics[width=8.0cm, height=7cm]{th12-th13_NH_d21_future.eps}
%\vglue -7.0cm \hglue 8.5cm
%\includegraphics[width=8.0cm, height=7cm]{th12-th13_IH_d31_future.eps}
%\caption{\label{fig:ovbb}
%}
%\end{figure}

\begin{figure}
\begin{center}
\includegraphics[width=16.0cm, height=6cm]{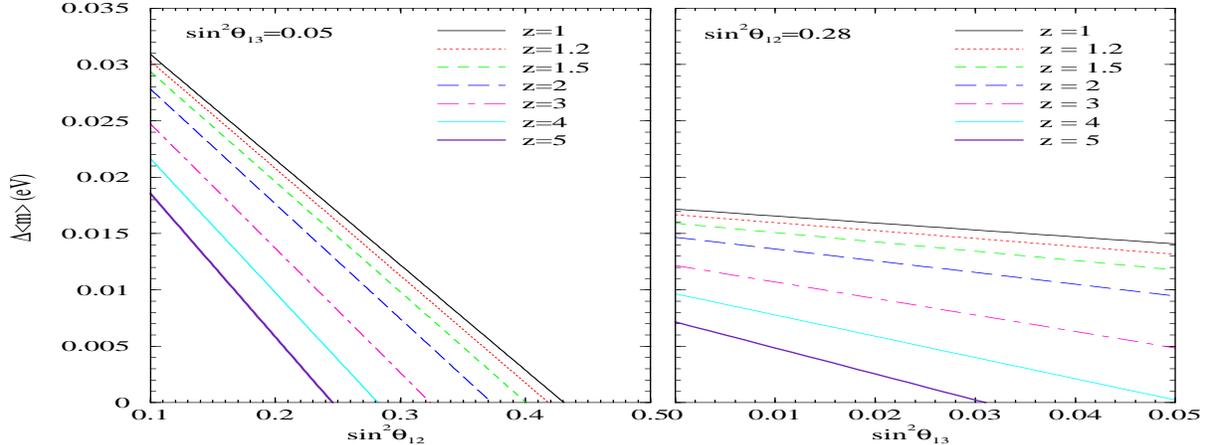}
\caption{\label{fig:diffnme}
The difference between the minimal value of \meff{} for IH 
and the maximal value of \meff{} for NH
for different $z$, 
as a function of $\sin^2 \theta_{12}$ 
(left--hand panel) and 
$\sin^2 \theta_{13}$ (right--hand panel).
}
\end{center}
\end{figure}

%\begin{figure}
%\begin{center}
%\includegraphics[width=14.0cm, height=7cm]{limitm0.eps}
%\caption{\label{fig:limitm0}
%}
%\end{center}
%\end{figure}

Arguably the most fundamental 
question, whether neutrinos are Dirac or Majorana particles, 
remains to be answered. If neutrinos are Majorana particles, 
in principle it should be possible to observe the process
$(A,Z) \rightarrow (A,Z-2) + 2 \, e^-~$, in which we have two 
$\beta$ particles in the final state and no neutrinos to accompany them
-- the process called neutrinoless double beta decay 
(\obb). The effective mass which 
will be extracted or bounded in a \obb{} experiment
is given by the following coherent sum: 
$\meff =  \left| \sum_i m_i \, U_{ei}^2 \right|$, 
where $m_i$ is the mass of the $i^{th}$ neutrino mass state, the sum 
is over all the light neutrino mass states and 
$U_{ei}$ are the matrix elements of the neutrino mixing matrix.
This means that \meff{} depends on 7 
out of 9 parameters contained in the neutrino mass matrix, 
not depending on only the Dirac phase $\delta$ and $\theta_{23}$. 
In particular, effective mass to be extracted from \onbb depends 
crucially on the neutrino mass spectrum. There have a been a 
large number of papers written on the implications 
of a future (non)measurement of \meff{} (see {\it e.g.}
\cite{rev_th}). 
The best current limit on the effective mass is given 
by the Heidelberg--Moscow collaboration \cite{HM} 
$\meff \le 0.35\, z~{\rm eV}$, 
where $z(={\cal O}(1))$ indicates that there is an uncertainty 
in the value of the \nme{} (NME) 
involved in the \obb{} process. 
Several new experiments are currently running, under 
construction or in the planing phase 
\cite{rev_ex}. Thus we expect \meff{} to be probed down to 
$\simeq 0.04$ eV and 
it would be pertinent to ask if such a measurement could help 
us learn about the neutrino mass hierarchy.

\noindent
For the NH scheme, for $m_1 \ll m_2 \ll m_3$ and 
assuming that $m_1$ can be neglected, we have
\be
\meffnh \simeq \left| \sqrt{\ms} \, \sss \, \csh + \sqrt{\ma} \, \sch \, 
e^{2i(\beta - \alpha)} 
\right|
\label{eq:meffnh}~.
\ee
For the IH scheme, assuming that 
$m_3 \ll m_1 < m_2$ and neglecting $m_3$, we have
\be
\meffih \simeq \sqrt{|\ma|} \, \csh \, \sqrt{ 1 - \sin^22\theta_{12} 
\, \sin^2
 \alpha} ~.
\label{eq:meffih}
\ee
Any future positive signal for $\obb$ will be able to distinguish
the IH scheme from the NH scheme 
if the difference between the predicted values 
for $\meff$ for the IH scheme and the NH scheme is larger than 
the error in the measured value of $\meff$. Among the most 
important errors involved is the one coming from the 
uncertainty involved in the value of the \nme. In Fig. 
\ref{fig:diffnme} we show the difference in the predicted 
values of $\meffnhmax$ and $\meffihmin$\footnote{Since 
$\meffnh$ and $\meffnh$ depend on the parameters of the neutrino 
mass matrix, they have a range of allowed values. 
$\meffnhmax$ and $\meffihmin$ are the largest and smallest 
predicted values for \meff{} in 
the NH and IH scheme respectively.},
taking into 
account the error in the \nme{}. This error is incorporated
through the parameter $z$, which gives the factor by which 
the \nme{} is uncertain. See \cite{ovbbus} for the details.
It was argued \cite{ovbbus} that for a given 
mass hierarchy, the uncertainty in the prediction of \meff{} 
coming from the uncertainty in the 
allowed values of $\ma$ and $\ms$
can be neglected, since these parameters are expected 
to be measured with very high accuracy in the immediate future.
Therefore the major uncertainty in \meff{} would come from 
the still uncertain range for the values of $\sss$ and $\sch$.
The Fig. \ref{fig:diffnme} shows the impact of the 
uncertainty in the values of $\sss$ and $\sch$ on the 
sensitivity of the future \obb{} experiments to the neutrino 
mass hierarchy. We see from the Fig. \ref{fig:diffnme} that for 
$\sch$ close to its current limit and assuming 
$z = 2$ and $\sss = 0.3$, we have 
$\Delta \meff \simeq 0.01$ eV and it should be possible to 
determine the neutrino hierarchy if  
the experimental uncertainty in \meff{} is less than 0.01 eV.
The chances of determining the hierarchy 
%improves as $\sch$ becomes smaller and 
is largest when $\sch=0$.
More importantly, we note that while the dependence on $\sch$ is 
weak, the sensitivity of the \obb{} experiments to the hierarchy
is strongly dependent on $\sss$. Therefore the uncertainty on 
$\sss$ should be reduced well enough to get the neutrino mass 
hierarchy using \obb{} experiments.

So far we assumed that the lightest neutrino mass was close to zero.
If the lightest neutrino had a mass $m_0 \gtap 0.01$ eV, 
it would not be possible to distinguish between the NH and IH schemes
using \obb{} measurements. 
%The mass spectrum in that case would be quasi-degenerate. 
However, we could still use \obb{} to put 
limit on the absolute neutrino mass scale. For quasi-degenerate (QD)
mass spectrum with a common mass scale $m_0$, the 
limit on the neutrino mass reads \cite{ovbbus} 
\be \label{eq:m0_lim}
m_0 \le z \, \meff_{\min}^{\rm exp} \,  \frac{1 + \tan^2 \theta_{12}}
{1 - \tan^2 \theta_{12} - 2 \, |U_{e3}|^2 } 
\equiv z \, \meff_{\min}^{\rm exp} \,  f(\theta_{12}, \theta_{13})~.
\ee 
Currently the uncertainty on $f(\theta_{12}, \theta_{13})$ is 
around 50\%, $1.9 < f(\theta_{12}, \theta_{13})< 5.6$.  
It is expected to reduce to $\sim$ 21\%($\sim$ 9)~\% at $3\sigma$ 
if a low energy $pp$ solar neutrino experiment
(reactor experiment at the SPMIN) 
would be built. The uncertainty depends only little on the value of 
$\theta_{13}$. 
From the current limit on the effective mass, 
$\meff \le 0.35 \, z$ eV, with the accepted value of $z \simeq 3$
and our current knowledge of 
$f(\theta_{12}, \theta_{13})$, we can set a limit on $m_0$ 
of 5.6 eV, clearly weaker than the limit from tritium beta 
decay experiments.
However,
if $f(\theta_{12}, \theta_{13})$ was known with an 
uncertainty of $20\%$, say 
$2.7 < f(\theta_{12}, \theta_{13})< 4.0$, then for 
$z \, \meff_{\min}^{\rm exp}=0.1$ eV  we could set the limits 
$0.3 ~{\rm eV} \ltap m_0 \ltap 0.4$ eV. 
Of course, if we have no signal for \obb, but just an upper limit 
on $z \, \meff_{\min}$, we have no longer an allowed range on $m_0$, 
but an upper limit corresponding to the largest value in the range. 
From the examples given above, one can note that 
for the QD mass spectrum, a measurement or 
a better constraint on \meff{} will 
lead to a stronger limit on the absolute neutrino mass 
scale compared to the current limit from 
direct kinematical searches.

%%%%%%%%%%%%%%%%%%%%%%%%%%%%%%%%%%%%%%%%%%%%%%%%%%%%%%%%
\section{Conclusions}
%%%%%%%%%%%%%%%%%%%%%%%%

A comparative study of the next generation solar neutrino 
experiments reveals that while $\ms$ can be measured 
well in the \kl experiment, to measure $\sss$ with a similar 
accuracy one would need a SPMIN reactor 
experiment with baseline of 
$60-70$ km. The forthcoming LBL experiments are expected to give 
very good limits on $\ma$ and $\sin^22\theta_{23}$. 
Comparable limits can be achieved using atmospheric
neutrino data in large magnetized iron detectors like ICAL.
The maximality of $\theta_{23}$ can be tested  
and the octant of $\theta_{23}$ can 
be determined most efficiently using matter effects in $\pmm$.
Neutrino mass hieararchy could be determined in either LBL 
experiments or in atmospheric $\numu/\anumu$ experiments if 
$\scht$ is large. In $\sch=0$, one could still see the hierarchy 
through \obb. We probed the conditions needed for it and 
pointed out that $\sss$ should be determined very accurately
to use a future measurement of \meff{} 
to determine the neutrino mass hierarchy. 

%\vskip 0.5cm
{\small 
I wish to thank my collaborators, A. Bandyopadhyay, 
S. Goswami, S. T. Petcov, W. Rodejohann, D. P. Roy and P. Roy.
It is also my pleasure to thank the organisers of NANP'05.}

%%%%%%%%%%%%%%%%%%%%%%%%%%

%%%%%%%%%%%%%%%%%%%%%%%%%%%%


\begin{thebibliography}{}
%%%%%%%%%%%%%%%%%%%%%%%%%%

\bibitem{skatm}
  Y.~Ashie {\it et al.},  
%[Super-Kamiokande Collaboration],
  %``A measurement of atmospheric neutrino oscillation parameters by
  %Super-Kamiokande I,''
  Phys.\ Rev.\ D {\bf 71}, 112005 (2005);
%  [arXiv:hep-ex/0501064].
  %%CITATION = HEP-EX 0501064;%%
%
  E.~Aliu {\it et al.},  
%[K2K Collaboration],
  %``Evidence for muon neutrino oscillation in an accelerator-based
  %experiment,''
  Phys.\ Rev.\ Lett.\  {\bf 94}, 081802 (2005).
 % [arXiv:hep-ex/0411038].
  %%CITATION = HEP-EX 0411038;%%

\bibitem{solar}
  T.~Araki {\it et al.},  
%[KamLAND Collaboration],
  %``Measurement of neutrino oscillation with KamLAND: Evidence of spectral
  %distortion,''
  Phys.\ Rev.\ Lett.\  {\bf 94}, 081801 (2005);
%  [arXiv:hep-ex/0406035].
  %%CITATION = HEP-EX 0406035;%%
%
  B.~Aharmim {\it et al.},  
%[SNO Collaboration],
  %``Electron energy spectra, fluxes, and day-night asymmetries of B-8 solar
  %neutrinos from the 391-day salt phase SNO data set,''
  arXiv:nucl-ex/0502021;
  %%CITATION = NUCL-EX 0502021;%%
%
S.~Fukuda {\it et al.},  
%[Super-Kamiokande Collaboration],
%``Determination of solar neutrino oscillation parameters using 1496 days  of Su per-Kamiokande-I data,''
Phys.\ Lett.\ B {\bf 539}, 179 (2002);
%[arXiv:hep-ex/0205075].
%%CITATION = HEP-EX 0205075;%%
%
%\cite{Ahmad:2002jz}
%\bibitem{Ahmad:2002jz}
%Q.~R.~Ahmad {\it et al.}  
%[SNO Collaboration],
%``Direct evidence for neutrino flavor transformation from neutral-current interactions in the Sudbury Neutrino Observatory,''
%Phys.\ Rev.\ Lett.\  {\bf 89}, 011301 (2002);
%[arXiv:nucl-ex/0204008];
%%CITATION = NUCL-EX 0204008;%%
%
%\cite{Ahmad:2002ka}
%\bibitem{Ahmad:2002ka}
%Q.~R.~Ahmad {\it et al.}  [SNO Collaboration],
%``Measurement of day and night neutrino energy spectra at SNO and  constraints on neutrino mixing parameters,''
%Phys.\ Rev.\ Lett.\  {\bf 89}, 011302 (2002);
%[arXiv:nucl-ex/0204009].
%%CITATION = NUCL-EX 0204009;%%
%
%\cite{Cleveland:nv}
%\bibitem{Cleveland:nv}
%\bibitem{cl}
B.~T.~Cleveland {\it et al.},
%``Measurement Of The Solar Electron Neutrino Flux With The Homestake  Chlorine Detector,''
Astrophys.\ J.\  {\bf 496}, 505 (1998);
%%CITATION = ASJOA,496,505;%%
%
%\bibitem{ga}
%\bibitem{Abdurashitov:2002nt}
%J.~N.~Abdurashitov {\it et al.},  
%[SAGE Collaboration],
% ``Measurement of the solar neutrino capture rate by the Russian-American
%gallium solar neutrino experiment during one half of the 22-year cycle  of
%solar activity,''
%J.\ Exp.\ Theor.\ Phys.\  {\bf 95}, 181 (2002);
%[Zh.\ Eksp.\ Teor.\ Fiz.\  {\bf 122}, 211 (2002)];
%[arXiv:astro-ph/0204245].
%%CITATION = ASTRO-PH 0204245;%%
%
%\cite{Hampel:1998xg}
%\bibitem{Hampel:1998xg}
%W.~Hampel {\it et al.},  
%[GALLEX Collaboration],
%``GALLEX solar neutrino observations: Results for GALLEX IV,''
%Phys.\ Lett.\ B {\bf 447}, 127 (1999);
%%CITATION = PHLTA,B447,127;%%
C. Cattadori, Talk at Neutrino 2004, Paris, France, June 14-19, 2004.

\bibitem{global}
  A.~Bandyopadhyay {\it et al.},
%, S.~Choubey, S.~Goswami, S.~T.~Petcov and D.~P.~Roy,
  %``Update of the solar neutrino oscillation analysis with the 766-Ty  KamLAND
  %spectrum,''
  Phys.\ Lett.\ B {\bf 608}, 115 (2005);
%  [arXiv:hep-ph/0406328].
  %%CITATION = HEP-PH 0406328;%%
%
  S.~Goswami, A.~Bandyopadhyay and S.~Choubey,
  %``Global analysis of neutrino oscillation,''
  Nucl.\ Phys.\ Proc.\ Suppl.\  {\bf 143}, 121 (2005).
%  [arXiv:hep-ph/0409224].
  %%CITATION = HEP-PH 0409224;%%


\bibitem{th12}
%\cite{Bandyopadhyay:2003du}
%\bibitem{Bandyopadhyay:2003du}
A.~Bandyopadhyay, S.~Choubey and S.~Goswami,
%``Exploring the sensitivity of current and future experiments to  Theta(odot),''
Phys.\ Rev.\ D {\bf 67}, 113011 (2003).
%[arXiv:hep-ph/0302243].
%%CITATION = HEP-PH 0302243;%%

\bibitem{shika}
  A.~Bandyopadhyay {\it et al.},
%, S.~Choubey, S.~Goswami and S.~T.~Petcov,
%``On the measurement of solar neutrino oscillation parameters with KamLAND,''
Phys.\ Lett.\ B {\bf 581}, 62 (2004).
%[arXiv:hep-ph/0309236].
%%CITATION = HEP-PH 0309236;%%

\bibitem{lownu} 
Y. Suzuki, talk at Neutrino 2004, June 14-19 (2004), Paris, France.
%S. Sch\"{o}nert,
%talk at Neutrino 2002, Munich, Germany,
%({\it http://neutrino2002.ph.tum.de}).

\bibitem{th12new}
  A.~Bandyopadhyay {\it et al.},
%, S.~Choubey, S.~Goswami and S.~T.~Petcov,
  %``High precision measurements of Theta(solar) in solar and reactor  neutrino
  %experiments,''
  Phys.\ Rev.\ D {\bf 72}, 033013 (2005).
%  [arXiv:hep-ph/0410283].
  %%CITATION = HEP-PH 0410283;%%

\bibitem{gdpaper}
%\cite{Beacom:2003nk}
%\bibitem{Beacom:2003nk}
J.~F.~Beacom and M.~R.~Vagins,
%``GADZOOKS! Antineutrino spectroscopy with large water Cherenkov detectors,''
arXiv:hep-ph/0309300.
%%CITATION = HEP-PH 0309300;%%

\bibitem{skgd}
%\cite{Choubey:2004bf}
%\bibitem{Choubey:2004bf}
S.~Choubey and S.~T.~Petcov,
%``Reactor anti-neutrino oscillations and gadolinium loaded Super-Kamiokande
%detector,''
Phys.\ Lett.\ B {\bf 594}, 333 (2004).
%[arXiv:hep-ph/0404103].
%%CITATION = HEP-PH 0404103;%%

\bibitem{maltoniglobal}
M.~Maltoni, {\it et al.}, 
%T.~Schwetz, M.~A.~Tortola and J.~W.~F.~Valle,
%``Status of global fits to neutrino oscillations,''
New J.\ Phys.\  {\bf 6}, 122 (2004), hep-ph/0405172 v7.

\bibitem{huber10}
%\bibitem{Huber:2004ug}
  P.~Huber {\it et al.},
%, M.~Lindner, M.~Rolinec, T.~Schwetz and W.~Winter,
  %``Prospects of accelerator and reactor neutrino oscillation experiments  for
  %the coming ten years,''
  Phys.\ Rev.\ D {\bf 70}, 073014 (2004).
%  [arXiv:hep-ph/0403068].
  %%CITATION = HEP-PH 0403068;%%

\bibitem{maltonimax23}
%\bibitem{Gonzalez-Garcia:2004cu}
  M.~C.~Gonzalez-Garcia, M.~Maltoni and A.~Y.~Smirnov,
  %``Measuring the deviation of the 2-3 lepton mixing from maximal with
  %atmospheric neutrinos,''
  Phys.\ Rev.\ D {\bf 70}, 093005 (2004).
%  [arXiv:hep-ph/0408170].
  %%CITATION = HEP-PH 0408170;%%

\bibitem{ino}
{\it http://www.imsc.res.in/$\sim$ino/}

\bibitem{inootherparams}
S.~Choubey and P.~Roy, in preparation.

\bibitem{antusch}
%\bibitem{Antusch:2004yx}
  S.~Antusch {\it et al.},
%, P.~Huber, J.~Kersten, T.~Schwetz and W.~Winter,
  %``Is there maximal mixing in the lepton sector?,''
  Phys.\ Rev.\ D {\bf 70}, 097302 (2004).
 % [arXiv:hep-ph/0404268].
  %%CITATION = HEP-PH 0404268;%%

\bibitem{inohier}
S.~Palomares-Ruiz and S.~T.~Petcov,
  %``Three-neutrino oscillations of atmospheric neutrinos, theta(13),  neutrino
  %mass hierarchy and iron magnetized detectors,''
  Nucl.\ Phys.\ B {\bf 712}, 392 (2005);
%\bibitem{Indumathi:2004kd}
  D.~Indumathi and M.~V.~N.~Murthy,
  %``A question of hierarchy: Matter effects with atmospheric neutrinos and
  %anti-neutrinos,''
  Phys.\ Rev.\ D {\bf 71}, 013001 (2005);
 % [arXiv:hep-ph/0407336].
  %%CITATION = HEP-PH 0407336;%%
  R.~Gandhi {\it et al.},
%, P.~Ghoshal, S.~Goswami, P.~Mehta and S.~Uma Sankar,
  %``Earth matter effects at very long baselines and the neutrino mass
  %hierarchy,''
  arXiv:hep-ph/0411252.
  %%CITATION = HEP-PH 0411252;%%


\bibitem{inomax}
S.~Choubey and P.~Roy,
arXiv:hep-ph/0509197.


\bibitem{minosmax}
  S.~Choubey and P.~Roy,
  %``Testing maximality in muon neutrino flavor mixing,''
  Phys.\ Rev.\ Lett.\  {\bf 93}, 021803 (2004). 

\bibitem{rev_th}
S.~T.~Petcov, 
%Invited talk given at the Nobel Symposium (N 129) on Neutrino Physics, 
%August 19 -- 24, 2004, Haga Slott, Enkoping, Sweden, 
  %``Theoretical Prospects of Neutrinoless Double Beta Decay,''
hep-ph/0504166 and references therein.
  %%CITATION = HEP-PH 0504166;%%

\bibitem{HM}H.~V.~Klapdor-Kleingrothaus {\it et al.},
  %``Latest results from the Heidelberg-Moscow double-beta-decay experiment,''
  Eur.\ Phys.\ J.\ A {\bf 12}, 147 (2001). 
%  [hep-ph/0103062].
  %%CITATION = HEP-PH 0103062;%%

\bibitem{rev_ex}C.~Aalseth {\it et al.},
  %``Neutrinoless double beta decay and direct searches for neutrino mass,''
  hep-ph/0412300 and references therein.
  %%CITATION = HEP-PH 0412300;%%

\bibitem{ovbbus}
  S.~Choubey and W.~Rodejohann,
  %``Neutrinoless double beta decay and future neutrino oscillation precision
  %experiments,''
  Phys.\ Rev.\ D {\bf 72}, 033016 (2005).
%  [arXiv:hep-ph/0506102].
  %%CITATION = HEP-PH 0506102;%%

%%%%%%%%%%%%%%%%%%%%%%%%%%%%
\end{thebibliography}
\end{document}